\documentclass[conference]{IEEEtran}
\IEEEoverridecommandlockouts
\usepackage[T1]{fontenc}%
\usepackage{verbatim}
\usepackage{cite}

\ifCLASSINFOpdf
\usepackage[pdftex]{graphicx}
\usepackage{tikz}
\usepackage{pgfplots}

\pgfplotsset{compat=newest}
\pgfplotsset{plot coordinates/math parser=false}
\newlength\figureheight

\else

\fi

\usepackage{amsmath}
\usepackage{mathrsfs}
\usepackage{amssymb}
\usepackage{amsthm}
\usepackage{thmtools}
\usepackage[ruled]{algorithm2e}
\usepackage{bm}
\usepackage{balance}
\usepackage{tabu}
\usepackage{longtable}
\newtheorem{example}{Example}
\newtheorem{theorem}{Theorem}

\newtheorem{definition}{Definition}

\newtheoremstyle{mydef}
	{3pt}		%
	{3pt}		%
	{}		%
	{}		%
	{\itshape}	%
	{:}		%
	{.5em}	%
	{}		%
\theoremstyle{mydef}

\newcommand{\Q}[0]{\mathcal{Q}}
\newcommand{\G}{\mathcal{G}}
\newcommand{\assign}[0]{\bm{P}}
\newcommand{\x}[0]{\bm{x}}

\newcommand{\A}[0]{\bm{A}}
\newcommand{\C}[0]{\bm{C}}
\newcommand{\T}[0]{\mathcal{T}}
\newcommand{\V}[0]{\mathcal{V}}

\newcommand{\W}[0]{\mathcal{W}}
\newcommand{\me}{\mathrm{e}}

\newcommand{\fexp}{L_{\text{BDC}}}

\renewcommand{\c}{\bm{c}}
\newcommand{\y}{\bm{y}}

\newcommand{\Z}{\mathcal{Z}}

\newcommand{\cmap}{\sigma_{\mathsf{map}}}
\newcommand{\cred}{\sigma_{\mathsf{reduce}}}

\newcommand{\tmap}{D_{\mathsf{map}}}

\newcommand{\tred}{D_{\mathsf{reduce}}}

\usepackage{color}
\newcommand{\todo}[1]{}
\renewcommand{\todo}[1]{{\color{red} TODO: {#1}}}

\ifCLASSOPTIONcompsoc
	\usepackage[caption=false,font=normalsize,labelfont=sf,textfont=sf]{subfig}
\else
	\usepackage[caption=false,font=footnotesize]{subfig}
\fi

\usepackage{algorithmic}
\usepackage[capitalize]{cleveref}
\crefname{equation}{\unskip}{\unskip}

\begin{document}

\title{Block-Diagonal Coding for Distributed Computing With Straggling Servers}

\author{
  \IEEEauthorblockN{
    Albin Severinson$^{\dag \, \ddag}$, Alexandre Graell i Amat$^\dag$, and Eirik
    Rosnes$^\ddag$}

  \IEEEauthorblockA{
    $\dag$Department of Electrical Engineering, Chalmers
    University of Technology, Gothenburg, Sweden}

  \IEEEauthorblockA{
    $\ddag$Simula@UiB, Bergen, Norway}

  \thanks{This work was funded by the Swedish Research Council under grant 2016-04253 and the
    Research Council of Norway under grant 240985/F20.}}

\maketitle

\begin{abstract}
  We consider the distributed computing problem of multiplying a set of vectors with a matrix. For
  this scenario, Li \emph{et al.} recently presented a unified coding framework and showed a
  fundamental tradeoff between computational delay and communication load. This coding framework is
  based on maximum distance separable (MDS) codes of code length proportional to the number of rows
  of the matrix, which can be very large. We propose a block-diagonal coding scheme consisting of
  partitioning the matrix into submatrices and encoding each submatrix using a shorter MDS code. We
  show that the assignment of coded matrix rows to servers to minimize the communication load can be
  formulated as an integer program with a nonlinear cost function, and propose an algorithm to solve
  it. We further prove that, up to a level of partitioning, the proposed scheme does not incur any
  loss in terms of computational delay (as defined by Li \emph{et al.}) and communication load
  compared to the scheme by Li \emph{et al.}. We also show numerically that, when the decoding time
  is also taken into account, the proposed scheme significantly lowers the overall computational
  delay with respect to the scheme by Li \emph{et al.}. For heavy partitioning, this is achieved at
  the expense of a slight increase in communication load.
\end{abstract}
\IEEEpeerreviewmaketitle

\section{Introduction}
Distributed computing has emerged as one of the most effective ways of tackling increasingly complex
computation problems. One of the main application areas is large-scale machine learning and data
analytics. Google routinely performs computations using several thousands of servers in their
MapReduce clusters \cite{Dean2004}. Distributed computing systems bring significant challenges.
Among them, the problems of straggling servers and bandwidth scarcity have recently received
significant attention. The straggler problem is a synchronization problem characterized by the fact
that a distributed computing task must wait for the slowest server to complete its computation. On
the other hand, distributed computing tasks typically require that data is moved between servers
during the computation, the so-called \textit{data shuffling}, which is a challenge in
bandwidth-constrained networks.

Coding for distributed computing to reduce the computational delay and the communication load
between servers has recently been considered in \cite{Li2015, Lee2015}. In
\cite{Li2015}, a structure of repeated computation tasks across servers was proposed, enabling
coded multicast opportunities that significantly reduce the required bandwidth to shuffle the
results. In \cite{Lee2015}, the authors showed that maximum distance separable
(MDS) codes can be applied to a linear computation task (e.g., multiplying a vector with a matrix)
to alleviate the effects of straggling servers and reduce the computational delay. In \cite{Li2016},
a unified coding framework was presented and a fundamental tradeoff between computational delay and
communication load was identified. The ideas of \cite{Li2015, Lee2015} can be
seen as particular instances of the framework in \cite{Li2016}, corresponding to the minimization of
the communication load or the computational delay.

The distributed computing problem addressed in \cite{Li2016} is a matrix multiplication problem
where a set of vectors $\x_1,\ldots,\x_N$ are multiplied by a matrix $\A$. In particular, as in
\cite{Lee2015}, the authors suggest the use of MDS codes, whose dimension is
equal to the number of rows of $\A$, to generate some redundant computations. In practice, the size
of $\A$ can be very large. For example, Google performs matrix-vector multiplications with matrices
of dimension of the order of $10^{10}\times 10^{10}$ when ranking the importance of websites
\cite{Ishii2014}. Since the decoding complexity of MDS codes on the packet erasure channel is
quadratic (for Reed-Solomon (RS) codes) in the code length \cite{Garr2013}, for very large matrix
sizes the decoding complexity may be prohibitively high.

In this paper, we introduce a block-diagonal encoding scheme for the distributed matrix
multiplication problem. The proposed encoding is equivalent to partitioning the matrix and applying
smaller MDS codes to each submatrix separately. The storage design for the proposed block-diagonal
encoding can be cast as an integer optimization problem with a nonlinear objective function, whose
computation scales exponentially with the problem size. We propose a heuristic solver for
efficiently solving the optimization problem, and a branch-and-bound approach for improving on the
resulting solution iteratively. We exploit a dynamic programming approach to speed up the
branch-and-bound operations. We prove that up to a certain partitioning level, partitioning does not
increase the computational delay (as defined in \cite{Li2016}) and the communication load with
respect to the scheme in \cite{Li2016}. Interestingly, when the decoding time is taken into account,
the proposed scheme achieves an overall computational delay significantly lower than the one of the
scheme in \cite{Li2016}. This is due to the fact that the proposed scheme allows using significantly
shorter MDS codes, hence reducing the decoding complexity and the decoding time. Also, numerical
results show that a high level of partitioning can be applied at the expense of only a slight
increase in the communication load.

\section{System Model}
\label{sec:SystemModel}

We consider the problem of multiplying a set of vectors with a matrix. In particular, given a matrix
$\A \in \mathbb{F}^{m \times n}$ and $N$ vectors $\x_1, \ldots, \x_N \in \mathbb{F}^{n}$, where $\mathbb{F}$ is some field, we want to
compute the $N$ vectors $\bm{y}_1 = \bm{Ax}_1, \ldots, \bm{y}_N = \bm{Ax}_N$. The computation is
performed in a distributed fashion using $K$ servers, $S_1,\ldots,S_K$, where each server stores
$\mu m$ matrix rows, for some $\frac{1}{K} \leq \mu \leq 1$. Prior to distributing the rows among
the servers, $\A$ is encoded by an $r\times m$ encoding matrix $\bm{\Psi}$, resulting in the coded
matrix $\bm{C} = \bm{\Psi \A}$, of size $r\times n$, i.e., the rows of $\A$ are encoded using an
$(r,m)$ linear code with $r \geq m$.

Let $q=K\frac{m}{r}$, where we assume that $r$ divides $Km$ and hence $q$ is an integer. The $r$
coded rows of $\C$, $\c_1,\ldots, \c_r$, are divided into $\binom{K}{\mu q}$ disjoint batches, each
containing $r/ \binom{K}{\mu q}$ coded rows. Each batch is assigned to $\mu q$ servers.
Correspondingly, a batch $B$ is labeled by a unique set $\T \subset \{S_1, \ldots, S_K\}$, of size
$|\T|=\mu q$, denoting the subset of servers that store that batch, and we write $B_\T$. Server
$S_k$, $k=1, \ldots, K$, stores the coded rows of $B_\T$ if and only if $S_k \in \T$.

\subsection{Probabilistic Runtime Model}
\label{sec:ProbRuntime}

We adopt the probabilistic model of the computation runtime of \cite{Lee2015}. We assume that
running a computation on a single server takes a random amount of time according to the shifted-exponential cumulative probability distribution \vspace{-1ex}
\begin{equation} \notag
  F(t) = \begin{cases}
    1 - \me^{-\left(\frac{t}{\sigma} - 1\right)}, & \text{for $t \geq \sigma$} \\
    0, & \text{otherwise}
\end{cases},
\end{equation}
where $\sigma$ is the number of multiplications and divisions required to complete the computation.
We do not take addition and subtraction into account as those operations are orders of magnitude
faster \cite{Zhi2001}.

When the computation is split into $K$ parallel subtasks running on separate servers, we assume
that the runtimes of the subtasks are independent and identically distributed random variables with
distribution $F(K t)$ \cite{Lee2015}. Furthermore, the runtime of the $g$-th, $g=1, \dots, K$,
fastest server to complete its subtask is given by the $g$-th order statistic, $F_{(g)}$, with
expectation \cite{Arnold2008} \vspace{-1ex}
\begin{equation} \notag
  \label{eq:runtime}
  f(\sigma, K, g) \triangleq \mathop{\mathbb{E}} \left(F_{(g)} \right) =
  \sigma \left(1 + \sum_{j=K-g+1}^K \frac{1}{j} \right).
\end{equation}

\subsection{Distributed Computing Model}
We consider the MapReduce framework described in \cite{Li2016}, where we assume that the input
vectors $\x_1, \ldots, \x_N$ are known to all servers. The overall computation then proceeds in
three phases: \textit{map}, \textit{shuffle}, and \textit{reduce}.

\subsubsection{Map Phase}
In the map phase, we compute in a distributed fashion coded intermediate values, which will be later
used to obtain vectors $\y_1,\ldots,\y_N$. Server $S$ multiplies the input vectors $\x_j$,
$j=1,\ldots,N$, by all the coded rows of matrix $\C$ it stores, i.e., it computes
\begin{equation} \notag
  \Z_{j}^{(S)} \triangleq \{\bm{c} \x_j : \c\in \{B_\T :\; S \in \T\}\},\; j=1, \ldots, N.
\end{equation}

The map phase terminates when a set of servers $\G \subseteq \{S_1, \ldots, S_K\}$ that collectively
store enough values to decode the output vectors have finished their map computations. We denote the
cardinality of $\G$ by $g$. The $(r,m)$ linear code proposed in \cite{Li2016} is an MDS code for
which $\y_1,\ldots,\y_N$ can be obtained from any subset of $q$ servers, i.e., $g = q$.

We define the computational delay of the map phase as its average runtime per source row and vector
$\y$, i.e., $\tmap = \frac{1}{mN} f(\frac{\cmap}{K}, K, g)$. $\tmap$ is referred to simply as the
computational delay in \cite{Li2016}. As all $K$ servers compute $\mu m$ inner products, each
requiring $n$ multiplications for each of the $N$ input vectors, we have $\cmap = K \mu m n N$.

After the map phase, the computation of $\y_1,\ldots,\y_N$ proceeds using only the servers in $\G$.
We denote by $\Q \subseteq \G$ the set of the first $q$ servers to complete the map phase. Each of
the $q$ servers in $\Q$ is responsible to compute $N/q$ of the vectors $\y_1,\ldots,\y_N$. Let
$\W_S$ be the set containing the indices of the vectors $\y_1,\ldots,\y_N$ server $S\in\Q$ is
responsible for. The remaining servers in $\G$ assist the servers in $\Q$ in the shuffle phase.

\subsubsection{Shuffle Phase} \label{sec:shuffle_phase}
In the shuffle phase, intermediate values calculated in the map phase are exchanged between servers
in $\G$ until all servers in $\Q$ hold enough values to compute the vectors they are responsible
for. As in \cite{Li2016}, we allow creating and multicasting coded messages that are simultaneously
useful for multiple servers. For a subset of servers $\mathcal{S} \subset \Q$ and
$S \in \Q \setminus \mathcal{S}$, we denote the set of intermediate values needed by server $S$ and
known \textit{exclusively} by the servers in $\mathcal{S}$ by $\V_{\mathcal{S}}^{(S)}$. More
formally,
\begin{equation} \notag
  \V_{\mathcal{S}}^{(S)} \triangleq \{ \bm{c} \x_j :\; j\in \W_S \text{ and } \c \in \{B_\T :\; \T \cap \Q =\mathcal{S} \} \}.
\end{equation}

Let $\alpha_j \triangleq \frac{\binom{q-1}{j} \binom{K-q}{\mu q - j}}{\frac{q}{K} \binom{K}{\mu q}}$
and $s_q \triangleq \text{inf}\left(s : \sum_{l=s}^{\mu q} \alpha_l \leq 1 - \mu \right)$. For each
$j \in \{ \mu q, \mu q - 1, \ldots, s_q\}$, and every subset $\mathcal{S} \subseteq \Q$ of size
$j + 1$, the shuffle phase proceeds as follows.

\begin{enumerate}
\item For each $S \in \mathcal{S}$, we evenly and arbitrarily split
  $\V_{\mathcal{S} \setminus S}^{(S)}$ into $j$ disjoint segments
  $\V_{\mathcal{S} \setminus S}^{(S)} = \{ \V_{\mathcal{S} \setminus S, \tilde{S}}^{(S)} : \tilde{S}
  \in \mathcal{S} \setminus S \}$, and associate the segment
  $\V_{\mathcal{S} \setminus S, \tilde{S}}^{(S)}$ with server
  $\tilde{S} \in \mathcal{S} \setminus S$.

\item Server $\tilde{S} \in \mathcal{S}$ multicasts the bit-wise XOR of all the segments associated
  with it in $\mathcal{S}$. More precisely, it multicasts
  $\mathop{\oplus}_{S \in \mathcal{S} \setminus \tilde{S}} \V_{\mathcal{S} \setminus
    S,\tilde{S}}^{(S)}$ to the other servers in $\mathcal{S} \setminus \tilde{S}$.
\end{enumerate}

For every pair of servers $S,\tilde{S}\in\mathcal{S}$, since server $S$ has computed locally the
segments $\V_{\mathcal{S} \setminus S', \tilde{S}}^{(S')}$ for all
$S' \in \mathcal{S} \setminus \{\tilde{S}, S\}$, it can cancel them from the message sent by server
$\tilde{S}$, and recover the intended segment. We finish the shuffle phase by either unicasting any
remaining needed values until all servers in $\Q$ hold enough intermediate values to decode
successfully, or by repeating the above two steps for $j=s_q - 1$, selecting the strategy achieving
the lower communication load.

\begin{definition}
  The \textit{communication load}, denoted by $L$, is the number of messages per source row and
  vector $\y$ exchanged during the shuffle phase, i.e., the total number of messages sent during the
  shuffle phase divided by $mN$.
\end{definition}

Depending on the strategy, the communication load after completing the multicast phase is
$\sum_{j=s_q}^{\mu q} \frac{\alpha_j}{j}$ or $\sum_{j=s_q-1}^{\mu q} \frac{\alpha_j}{j}$
\cite{Li2016}. For the scheme in \cite{Li2016}, the total communication load is

{\begin{small}
  \begin{equation} \label{eq:unpartitioned_load} L_{\text{MDS}} = \min\left( \sum_{j=s_q}^{\mu q}
      \frac{\alpha_j}{j} + 1 - \mu - \sum_{j=s_q}^{\mu q} \alpha_j, \sum_{j=s_q -1 }^{\mu q}
      \frac{\alpha_j}{j} \right).
  \end{equation}
\end{small}}

As in \cite{Li2016}, we consider the cost of a multicast message to be equal to that of a unicast
message. In real systems, however, it may vary depending on the network architecture.

\subsubsection{Reduce Phase}
Finally, in the reduce phase, the vectors $\y_1,\ldots,\y_N$ are computed. More specifically, server
$S\in \Q$ uses the locally computed sets $\Z_{1}^{(S)}, \ldots, \Z_{N}^{(S)}$ and the received
messages to compute the vectors in $\{\y_j:j\in\W_{S}\}$. The computational delay of the reduce
phase is its average runtime per source row and output vector $\y$, i.e.,
$\tred = \frac{1}{mN} f(\frac{\cred}{q}, q, q)$, where $\cred$ is given in
\cref{computational_delay}.
\begin{definition}
\label{def:D}
  The \textit{overall computational delay}, $D$, is the sum of the map and reduce phase delays, i.e.,
  $D = \tmap + \tred$.
\end{definition}

\section{Block-Diagonal Coding}

We introduce  a block-diagonal encoding matrix of the form
\begin{equation} \notag
  \bm{\Psi} =
  \begin{bmatrix}
    \bm{\psi}_1 & &  \\
    & \ddots & \\
    & & \bm{\psi}_T
  \end{bmatrix},
\end{equation}

\noindent
where $\bm{\psi}_1, \ldots, \bm{\psi}_T$ are $\frac{r}{T}\times \frac{m}{T}$ encoding matrices of an
$(\frac{r}{T},\frac{m}{T})$ MDS code, for some integer $T$ that divides $m$ and $r$. Note that the
encoding given by $\bm{\Psi}$ amounts to partitioning the rows of $\A$ into $T$ disjoint submatrices
$\A_1, \ldots, \A_T$ and encoding each submatrix separately. We refer to an encoding $\bm{\Psi}$
with $T$ disjoint submatrices as a $T$-partitioned scheme, and to the submatrix of $\C=\bm{\Psi}\A$
corresponding to $\bm{\psi}_i$ as the $i$-th partition. We remark that all submatrices can be
encoded using the same encoding matrix, i.e., $\bm{\psi}_i=\bm{\psi},\; i=1,\ldots,T$, reducing the
storage requirements, and encoding/decoding can be performed in parallel if many servers are
available. We further remark that the case $\bm{\Psi}=\bm{\psi}$ (i.e., the number of partitions is
$T=1$) corresponds to the scheme in \cite{Li2016}, which we will sometimes refer to as the
\emph{unpartitioned} scheme.

\subsection{Assignment of Coded Rows to Batches}
For a block-diagonal encoding matrix $\bm{\Psi}$, we denote by $\bm{c}_i^{(t)}$, $t=1,\ldots,T$ and
$i=1,\ldots,r/T$, the $i$-th coded row of $\C$ within partition $t$. For example, $\bm{c}_1^{(2)}$
denotes the first coded row of the second partition. As described in Section~\ref{sec:SystemModel},
the coded rows are divided into $\binom{K}{\mu q}$ disjoint batches. To formally describe the
assignment of coded rows to batches we use a $\binom{K}{\mu q} \times T$ integer matrix
$\assign=[p_{i,j}]$, where $p_{i,j}$ describes the number of rows from partition $j$ that are stored
in batch $i$. Note that, due to the MDS property, any set of $m/T$ rows of a partition is sufficient
to decode the partition. Thus, without loss of generality, we consider a \emph{sequential}
assignment of rows of a partition into batches. For example, for the assignment $\assign$ in
Example~\ref{ex:Example} (see \eqref{eq:storage_design}), rows $\c_1^{(1)}$ and $\c_2^{(1)}$ are
assigned to batch $1$, $\c_3^{(1)}$ and $\c_4^{(1)}$ are assigned to batch $2$, and so on. %
The rows of $\assign$ are labeled by the subset of servers the corresponding batch is stored at, and
the columns are labeled by its partition index. We refer to the pair $(\bm{\Psi},\assign)$ as the
\textit{storage design}. The assignment matrix $\bm P$ must satisfy the following conditions.
\begin{enumerate}
\item The entries of each row of $\assign$ must sum to the batch size, i.e.,
  $\sum_{j=1}^T p_{i, j} = r/\binom{K}{\mu q},\; 1 \leq i \leq \binom{K}{\mu q}$.
\item The entries of each column of $\assign$ must sum to the number of rows per partition, i.e.,
  $\sum_{i=1}^{\binom{K}{\mu q}} p_{i, j} = \frac{r}{T},\; 1 \leq j \leq T$.

\end{enumerate}

\begin{example}[$m=20$, $N=4$, $K=6$, $q=4$, $\mu=1/2$, $T=5$]
  \begin{figure}[!t]
    \centering
    \resizebox{0.8\columnwidth}{!}{


\tikzstyle{block} = [rectangle, rounded corners, minimum width=3cm, minimum height=1cm,text centered, thick, draw=black]

\tikzstyle{arrow} = [->,>=stealth',font=\scriptsize,rounded corners]

\newcommand{\shift}{0.28cm}
\newcommand{\bigshift}{2cm}
\newcommand{\itsstackwidth}{1cm}

\begin{tikzpicture}[node distance=2cm, block/.append style={minimum width=2cm}]

  \node [block, node distance=3cm, text width=3.3cm] (server1)
  {$\bm{c}_1^{(1)}$ $\bm{c}_3^{(1)}$ $\bm{c}_5^{(1)}$ $\bm{c}_1^{(2)}$ $\bm{c}_3^{(2)}$ \\ $\bm{c}_2^{(1)}$ $\bm{c}_4^{(1)}$ $\bm{c}_6^{(1)}$
    $\bm{c}_2^{(2)}$ $\bm{c}_4^{(2)}$ \\ \textbf{Server} $\bm{S_1}$};

  \node [block, right of=server1, node distance=3.6cm, text width=3.3cm] (server2)
  {$\bm{c}_1^{(1)}$ $\bm{c}_5^{(2)}$ $\bm{c}_1^{(3)}$ $\bm{c}_3^{(3)}$ $\bm{c}_5^{(3)}$ \\ $\bm{c}_2^{(1)}$ $\bm{c}_6^{(2)}$ $\bm{c}_2^{(3)}$
    $\bm{c}_4^{(3)}$ $\bm{c}_6^{(3)}$ \\ \textbf{Server} $\bm{S_2}$};

  \node [block, right of=server2, node distance=3.6cm, text width=3.3cm] (server3)
  {$\bm{c}_3^{(1)}$ $\bm{c}_5^{(2)}$ $\bm{c}_1^{(4)}$ $\bm{c}_3^{(4)}$ $\bm{c}_5^{(4)}$ \\ $\bm{c}_4^{(1)}$ $\bm{c}_6^{(2)}$ $\bm{c}_2^{(4)}$
    $\bm{c}_4^{(4)}$ $\bm{c}_6^{(4)}$ \\ \textbf{Server} $\bm{S_3}$};

  \node [block, below of=server1, node distance=1.6cm, text width=3.3cm] (server4)
  {$\bm{c}_5^{(1)}$ $\bm{c}_1^{(3)}$ $\bm{c}_1^{(4)}$ $\bm{c}_1^{(5)}$ $\bm{c}_3^{(5)}$ \\ $\bm{c}_6^{(1)}$ $\bm{c}_2^{(3)}$ $\bm{c}_2^{(4)}$
    $\bm{c}_2^{(5)}$ $\bm{c}_4^{(5)}$ \\ \textbf{Server} $\bm{S_4}$};

  \node [block, below of=server2, node distance=1.6cm, text width=3.3cm] (server5)
  {$\bm{c}_1^{(2)}$ $\bm{c}_3^{(3)}$ $\bm{c}_3^{(4)}$ $\bm{c}_1^{(5)}$ $\bm{c}_5^{(5)}$ \\ $\bm{c}_2^{(2)}$ $\bm{c}_4^{(3)}$ $\bm{c}_4^{(4)}$
    $\bm{c}_2^{(5)}$ $\bm{c}_6^{(5)}$ \\ \textbf{Server} $\bm{S_5}$};

  \node [block, below of=server3, node distance=1.6cm, text width=3.3cm] (server6)
  {$\bm{c}_3^{(2)}$ $\bm{c}_5^{(3)}$ $\bm{c}_5^{(4)}$ $\bm{c}_3^{(5)}$ $\bm{c}_5^{(5)}$ \\ $\bm{c}_4^{(2)}$ $\bm{c}_6^{(3)}$ $\bm{c}_6^{(4)}$
    $\bm{c}_4^{(5)}$ $\bm{c}_6^{(5)}$ \\ \textbf{Server} $\bm{S_6}$};

\end{tikzpicture}
    }
    \vspace{-2ex}
    \caption{Storage design for $m=20$, $N=4$, $K=6$, $q=4$,
      $\mu=1/2$, and $T=5$.}
    \label{fig:example_storage}
    \vspace{-3ex}
  \end{figure}

  \label{ex:Example}
  \begin{figure}[!t]
    \centering
    \resizebox{0.8\columnwidth}{!}{


\tikzstyle{block} = [rectangle, rounded corners, minimum width=3cm, minimum height=1cm,text centered, thick, draw=black]
\tikzstyle{arrow} = [->,font=\scriptsize,rounded corners, thick]

\newcommand{\shift}{0.28cm}
\newcommand{\bigshift}{2cm}
\newcommand{\itsstackwidth}{1cm}

\begin{tikzpicture}[node distance=2cm, block/.append style={minimum width=2cm}]

  \node [block, node distance=3cm, text width=3.3cm] (server1)
  {$\bm{c}_1^{(1)}$ $\bm{c}_3^{(1)}$ $\bm{c}_5^{(1)}$ $\bm{c}_1^{(2)}$ $\bm{c}_3^{(2)}$ \\ $\bm{c}_2^{(1)}$ $\bm{c}_4^{(1)}$ $\bm{c}_6^{(1)}$
    $\bm{c}_2^{(2)}$ $\bm{c}_4^{(2)}$ \\ \textbf{Server} $\bm{S_1}$};

  \node [below of=server1, node distance=1.1cm, text width=3cm, text centered] (multicast1)
  {$\bm{c}_1^{(1)} \x_3 \mathop{\oplus} \bm{c}_3^{(1)} \x_2$};

  \node [block, below of=server1, node distance=10cm, left
  of=server1, node distance=2.8cm, text width=3.3cm] (server2)
  {$\bm{c}_1^{(1)}$ $\bm{c}_5^{(2)}$ $\bm{c}_1^{(3)}$ $\bm{c}_3^{(3)}$ $\bm{c}_5^{(3)}$ \\ $\bm{c}_2^{(1)}$ $\bm{c}_6^{(2)}$ $\bm{c}_2^{(3)}$
    $\bm{c}_4^{(3)}$ $\bm{c}_6^{(3)}$ \\ \textbf{Server} $\bm{S_2}$};

  \node [above of=server2, node distance=1.1cm, text width=3cm, text centered] (multicast2)
  {$\bm{c}_2^{(1)} \x_3 \mathop{\oplus} \bm{c}_5^{(2)} \x_1$};

  \node [block, below of=server1, node distance=2.4cm, right
  of=server1, node distance=2.8cm, text width=3.3cm] (server3)
  {$\bm{c}_3^{(1)}$ $\bm{c}_5^{(2)}$ $\bm{c}_1^{(4)}$ $\bm{c}_3^{(4)}$ $\bm{c}_5^{(4)}$ \\ $\bm{c}_4^{(1)}$ $\bm{c}_6^{(2)}$ $\bm{c}_2^{(4)}$
    $\bm{c}_4^{(4)}$ $\bm{c}_6^{(4)}$ \\ \textbf{Server} $\bm{S_3}$};

  \node [above of=server3, node distance=1.1cm, text width=3cm, text centered] (multicast3)
  {$\bm{c}_4^{(1)} \x_2 \mathop{\oplus} \bm{c}_6^{(2)} \x_1$};

  \draw [arrow, transform canvas={xshift=1ex, yshift=-1ex}]
  (multicast1.west) -- (multicast2.north);

  \draw [arrow, transform canvas={xshift=-1ex, yshift=-1ex}]
  (multicast1.east) -- (multicast3.north);

  \draw [arrow, transform canvas={xshift=-1ex, yshift=1ex}]
  (multicast2.north) -- (multicast1.west);

  \draw [arrow, transform canvas={xshift=1ex, yshift=1ex}]
  (multicast3.north) -- (multicast1.east);

  \draw [arrow, transform canvas={xshift=-1ex}]
  (multicast2.east) -- (server3.west);

  \draw [arrow, transform canvas={xshift=1ex}]
  (multicast3.west) -- (server2.east);

\end{tikzpicture}
    }
    \vspace{-2ex}
    \caption{Multicasting coded values between servers $S_1$, $S_2$, and $S_3$.}
    \label{fig:example_shuffling}
    \vspace{-3ex}
  \end{figure}

  For these parameters, there are $r/T=6$ coded rows per partition, of which $m/T=4$ are sufficient
  for decoding, and $\binom{K}{\mu q}=15$ batches, each containing $r/ \binom{K}{\mu q}=2$ coded
  rows. We construct the storage design shown in \cref{fig:example_storage} with assignment matrix
  {\small
\begin{equation}\label{eq:storage_design}
  \assign = \bordermatrix{
    ~ & 1 & 2 & 3 & 4 & 5 \cr
    (S_1, S_2) & 2 & 0 & 0 & 0 & 0 \cr
    (S_1, S_3) & 2 & 0 & 0 & 0 & 0 \cr
    (S_1, S_4) & 2 & 0 & 0 & 0 & 0 \cr
    (S_1, S_5) & 0 & 2 & 0 & 0 & 0 \cr
    ~~~~\vdots & & & \vdots & & \cr
    (S_4, S_6) & 0 & 0 & 0 & 0 & 2 \cr
    (S_5, S_6) & 0 & 0 & 0 & 0 & 2 \cr},
\end{equation}}%
\noindent
where rows are labeled by the subset of servers the batch is stored at, and columns are labeled by
the partition index. For this storage design, any $g=4$ servers collectively store at least 4 coded
rows from each partition. However, some servers store more rows than needed to decode some
partitions, suggesting that this storage design is suboptimal.

Assume $\G=\{S_1, S_2, S_3, S_4\}$ is the set of $g=4$ servers that finish their map computations
first. Also, assign vector $\y_i$ to server $S_i$, $i=1,2,3,4$. We illustrate the coded shuffling
scheme for $\mathcal{S} = \{S_1, S_2, S_3\}$ in \cref{fig:example_shuffling}. $S_1$ multicasts
$\bm{c}_1^{(1)} \x_3 \mathop{\oplus} \bm{c}_3^{(1)} \x_2$ to $S_2$ and $S_3$. Since $S_2$ and $S_3$
can cancel $\bm{c}_1^{(1)} \x_3$ and $\bm{c}_3^{(1)} \x_2$, respectively, both servers receive one
needed intermediate value. Similarly, $S_2$ multicasts
$\bm{c}_2^{(1)} \x_3 \mathop{\oplus} \bm{c}_5^{(2)} \x_1$, while $S_3$ multicasts
$\bm{c}_4^{(1)} \x_2 \mathop{\oplus} \bm{c}_6^{(2)} \x_1$. This process is repeated for
$\mathcal{S} = \{S_2, S_3, S_4\}$, $\mathcal{S} = \{S_1, S_3, S_4\}$, and
$\mathcal{S} = \{S_1, S_2, S_4\}$. After the shuffle phase, we have sent $12$ multicast messages and
$30$ unicast messages, resulting in a communication load of $(12 + 30)/20/4=0.525$, a $50\%$
increase from the load of the unpartitioned scheme ($0.35$, given by \cref{eq:unpartitioned_load}).
In this case, $S_1$ received additional intermediate values from partition $2$, despite already
storing enough, further indicating that the assignment in \eqref{eq:storage_design} is suboptimal.

\end{example}

\section{Performance of the Block-Diagonal Coding}
In this section, we analyze the performance impact of partitioning. We have the following theorem.
\begin{theorem}
  For $T \leq r/\binom{K}{\mu q}$ there exists an assignment matrix $\assign$ such that the
  communication load and computational delay of the map phase are equal to those of the
  unpartitioned scheme.
\end{theorem}

\subsection{Communication Load}
For the unpartitioned scheme of \cite{Li2016}, $\G = \Q$, and the number of remaining values that
need to be unicasted after the multicast phase is constant, regardless which subset $\Q$ of servers
finish first their map computations. However, for the block-diagonal (partitioned) coding scheme,
both $g$ and the number of remaining unicasts may vary.

For a given assignment $\assign$ and a specific $\Q$, we denote by $U_\Q^{(S)}(\assign)$ the number of remaining values
needed after the multicast phase by server $S \in \Q$, and by
$U_\Q(\assign) \triangleq \sum_{S \in \Q} U_\Q^{(S)}(\assign)$ the total
number of remaining values needed by the servers in $\Q$. Let $\mathbb{Q}^q$ denote the
superset of all sets $\Q$ and define
$L_{\mathbb{Q}} \triangleq \frac{1}{mN} \frac{1}{\left\vert \mathbb{Q}^q \right\vert} \sum_{\Q \in
  \mathbb{Q}^q} U_\Q(\assign)$. Then, for a given storage design $(\bm{\Psi},\assign)$, the
communication load of the block-diagonal coding scheme is given by
\begin{small}
\begin{equation} \label{eq:objective_function_expected}
    \fexp(\bm{\Psi}, \assign) =
    \min\left( \sum_{j=s_q}^{\mu q} \frac{\alpha_j}{j} + L_{\mathbb{Q}},
      \sum_{j=s_q-1}^{\mu q} \frac{\alpha_j}{j} + L_{\mathbb{Q}} \right),
  \end{equation}%
\end{small}%
where $L_{\mathbb{Q}}$ depends on the shuffling scheme (see
\cref{sec:shuffle_phase}) and is different in the first and second term of the
minimization in \eqref{eq:objective_function_expected}. To evaluate
$U_\Q^{(S)}$, we count the total number of intermediate values that need to be
unicasted to server $S$ until it holds $m/T$ intermediate values from each
partition.

For a given $\bm{\Psi}$, the assignment of rows into batches can be formulated as an optimization
problem, where one would like to minimize $\fexp$ over all assignments $\assign$. More precisely,
the optimization problem is $\min_{\assign \in \mathbb{P}} \fexp(\bm{\Psi},\assign)$, where
$\mathbb{P}$ is the set of all assignments $\assign$, and where the dependence of $\fexp$ on
$\assign$ is nonlinear. This is a computationally complex problem since both the complexity of
evaluating the performance of a given assignment and the number of assignments scale exponentially
in the problem size.

\subsection{Computational Delay}
\label{computational_delay}

We consider the delay incurred by both the map and reduce phases (see Definition~\ref{def:D}). Note
that in \cite{Li2016} only $\tmap$ is considered, i.e., $D = \tmap$. However, one should not neglect
the computational delay incurred by the reduce phase. Thus, one should consider $D = \tmap + \tred$.
The reduce phase consists of decoding the $N$ output vectors and hence the delay it incurs depends
on the code and decoding algorithm. We assume that each partition is encoded using an RS code and is
decoded using the Berlekamp-Massey algorithm. We measure the decoding complexity by its associated
shifted-exponential parameter $\sigma$ (see \cref{sec:ProbRuntime}).

The number of field multiplications required to decode an $(r/T, m/T$) RS code is $(r / T)^2 \xi$
\cite{Zhi2001}, where $\xi$ is the fraction of erased symbols. With $\xi$ upperbounded by
$1 - \frac{q}{K}$ (the map phase terminates when a fraction of at least $\frac{q}{K}$ symbols from
each partition is available) the complexity of decoding the $T$ partitions for all $N$ output
vectors is at most
\begin{equation}
  \label{eq:decoding_complexity}
  \cred = \frac{r^2 (1 - \frac{q}{K}) N}{T}.
\end{equation}
The decoding complexity of the scheme in \cite{Li2016} is given by evaluating
\cref{eq:decoding_complexity} for $T=1$. By choosing $T$ close to $r$, we can thus significantly
reduce the delay of the reduce phase.

\section{Assignment Solvers}
We propose two solvers for the problem of assigning rows into batches: a heuristic solver that is
fast even for large problem instances, and a hybrid solver combining the heuristic solver with a
branch-and-bound solver. The branch-and-bound solver produces an optimal assignment but is
significantly slower, hence it can be used as stand-alone only for small problem instances. We use a
dynamic programming approach to speed up the branch-and-bound solver by caching $U_\Q^{(S)}$ for all
$\Q \in \mathbb{Q}^q$, indexed by the batches each $U_\Q^{(S)}$ is computed from. This way we
only need to update the affected $U_\Q^{(S)}$ when assigning a row to a batch. For all solvers, we
first label the batches lexiographically and then optimize $\fexp$ in
\cref{eq:objective_function_expected}. The solvers are available under the Apache 2.0 license
\cite{severinson17}.

\subsection{Heuristic Solver}
The heuristic solver creates an assignment matrix $\assign$ in two steps. We first set each entry of
$\assign$ to $\gamma \triangleq \left\lfloor r/\left(\binom{K}{\mu q}\cdot T\right)\right\rfloor$,
thus assigning the first $\binom{K}{\mu q}\gamma$ rows of each partition to batches such that each
batch is assigned $\gamma T$ rows. Let $d=r/\binom{K}{\mu q}-\gamma T$ be the number of rows that
still need to be assigned to each batch. The $r/T-\binom{K}{\mu q}\gamma$ rows per partition not
assigned yet are assigned in the second step as given in the algorithm below. \vspace{-2ex}
\begin{algorithm}[!h]
\SetKwInOut{Input}{Input}
\SetKwInOut{Output}{Output}
\Input{$\bm P$, $d$, $K$, $T$, and $\mu q$}
\For{$0\le a < d\binom{K}{\mu q}$}{\label{algcPoP:outerWhile}
	$i \leftarrow \lfloor a/d \rfloor +1$\\
	$j \leftarrow (a \bmod T)+1$\\
	$p_{i,j} \leftarrow p_{i,j}+1$
			}
\KwRet{$\bm P$}
\end{algorithm}

\vspace{-3ex}

\subsection{Branch-and-Bound Solver} The branch-and-bound solver finds an optimal solution by
recursively branching at each batch for which there is more than one possible assignment and
considering all options. For each branch, we lowerbound the value of the objective function of any
assignment in that branch and only investigate branches with possibly better assignments.

\subsubsection{Branch}For the first row of $\assign$ with remaining assignments, branch
on every available assignment for that row.

\subsubsection{Bound} We keep a record of all nonzero $U_\Q^{(S)}$ for all $\Q$ and $S$, and index
them by the batches they are computed from. An assignment to a batch can at most reduce $\fexp$ by
$1 / \left(m N \left\vert\mathbb{Q}^q\right\vert \right)$ for each nonzero $U_\Q^{(S)}$ indexed by
that batch, and we lowerbound $\fexp$ for a subtree by assuming that no $U_\Q^{(S)}$ will drop to
zero for any subsequent assignment.

\subsection{Hybrid Solver} We first find a candidate solution using the heuristic solver and then
iteratively improve it using the branch-and-bound solver. In particular, we decrement by $1$ a
random set of elements of $\assign$ and then use the branch-and-bound solver to reassign the
corresponding rows optimally. We repeat this process until the average improvement between
iterations drops below some threshold.

\section{Numerical Results}
\begin{figure}[!t]
  \centering
  \resizebox{\columnwidth}{!}{
    \input{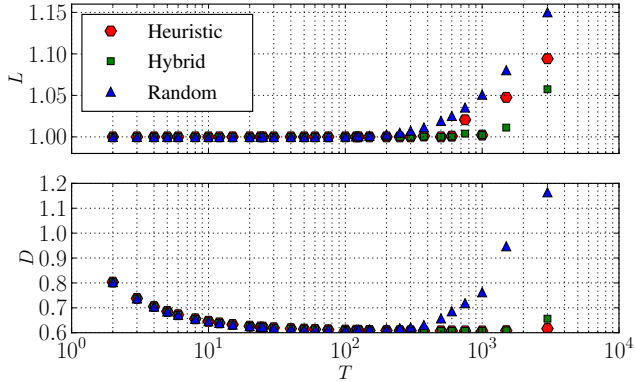}
  }
  \vspace{-6ex}
  \caption{The tradeoff between partitioning and performance for $m=6000$, $n=6000$, $K=9$, $q=6$,
    $N=6$, and $\mu=1/3$.}
  \label{fig:partitioning_comparison}
  \vspace{-4ex}
\end{figure}

\begin{figure}[!t]
  \centering
  \resizebox{\columnwidth}{!}{
    \input{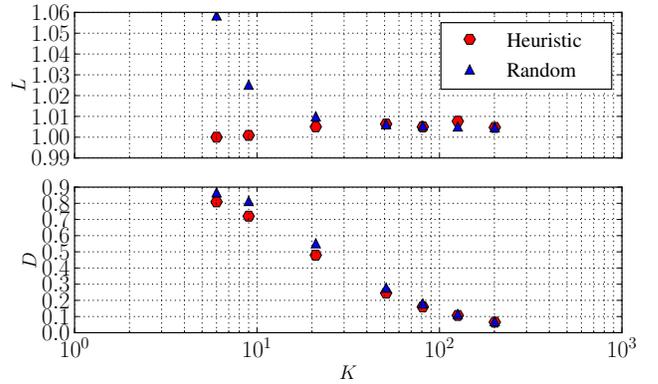}
  }
  \vspace{-6ex}
  \caption{Performance dependence on system size for $\mu q=2$, $n=10000$, $\mu m=2000$, $m/T=10$
    rows per partition, and code rate $m/r=2/3$.}
  \label{fig:system_size_comparison}
  \vspace{-4ex}
\end{figure}

In Figs.~\ref{fig:partitioning_comparison} and~\ref{fig:system_size_comparison}, we plot the
performance of the proposed block-diagonal coding scheme with assignment $\assign$ given by the
heuristic and the hybrid solvers. We normalize the performance by that of the unpartitioned scheme
of \cite{Li2016}. We also give the average performance over $100$ random assignments.

In \cref{fig:partitioning_comparison}, we plot the normalized communication load
($L$) and overall computational delay ($D$) as a function of the number of
partitions $T$. The system parameters are $m=6000$, $n=6000$, $K=9$, $q=6$,
$N=6$, and $\mu=1/3$. For up to $r / \binom{K}{\mu q}=250$ partitions, the
proposed scheme does not incur any loss in $\tmap$ and communication load with
respect to the unpartitioned scheme. However, the proposed scheme yields a
significantly lower $D$ compared to the scheme in \cite{Li2016} (about $40\%$
speedup for $T>50$). For heavy partitioning (around $T=800$) a tradeoff between
partitioning level, communication load, and map phase delay is observed. With
$3000$ partitions (the maximum possible), there is about a $10\%$ increase in
communication load. Note that the gain in overall computational delay saturates
with the partitioning level, thus there is no reason to partition beyond a given
level.

In \cref{fig:system_size_comparison}, we plot the normalized performance for a constant $\mu q = 2$,
$n=10000$, $\mu m = 2000$, $m/T=10$ rows per partition, and code rate $m/r=2/3$ as a function of
the number of servers, $K$. The results shown are averages over $1000$ randomly generated
realizations of $\G$ as it is computationally unfeasible to evaluate the performance exhaustively in
this case. The heuristic solver outperforms the random assignments for small $K$, but as $K$ grows
the performance of both solvers converge. The delay is an order of magnitude lower than that of the
scheme in \cite{Li2016} for the largest system considered.

\section{Conclusion}
We introduced a block-diagonal coding scheme for distributed matrix multiplication based on
partitioning the matrix into smaller submatrices. Compared to earlier (MDS) schemes, the proposed
scheme yields a significantly lower computational delay with no increase in communication load up to
a level of partitioning. For instance, for a square matrix with $6000$ rows and columns, the
proposed scheme reduces the computational delay by about $40\%$ when the number of partitions
$T>50$. We have also considered fountain codes, and will present the results in an extension of this
paper.

\ifCLASSOPTIONcaptionsoff
  \newpage
\fi

\bibliographystyle{IEEEtran}
\bibliography{manuscript}{}

\end{document}